\documentclass[10pt, twocolumn, letterpaper, journal]{IEEEtran}

\input packages-headers

\RequirePackage[top=2.0cm,bottom=2.0cm,left=1.5cm,right=1.5cm,nohead,nofoot]{geometry}

\newcommand{\Q}{$Q$}

\begin{document}

\ignore{ 
    \definecolor{mylightgrey}{rgb}{0.6,0.6,0.6}
    \definecolor{mydarkgrey}{rgb}{0.3,0.3,0.3}
    \definecolor{mybrown}{rgb}{0.6,0.2,0}
    \definecolor{mylightgreen}{rgb}{0,1,0}
    \definecolor{mydarkgreen}{rgb}{0,0.5,0}
    \definecolor{myorange}{rgb}{1,0.4,0}
    \definecolor{mypurple}{rgb}{0.8,0,0.8}
}

\renewcommand{\refname}{References}


\let\redHL=\ignore
\let\jnote=\ignore
\let\tnote=\ignore
\let\tsumm=\ignore

\pagestyle{empty}

\title{{\huge
Rigorous \Q{} Factor Formulation and Characterization for Nonlinear Oscillators}
}

\author
{
Tianshi Wang and Jaijeet Roychowdhury\\
{\normalsize
The Department of Electrical Engineering and Computer Sciences, The University of California, Berkeley, CA, USA}\\
Email: \texttt{\{tianshi, jr\}@berkeley.edu}
\vspace{-1em}
}

\maketitle

\begin{abstract}
In this paper, we discuss the definition of \Q{} factor for nonlinear
oscillators.
While available definitions of \Q{} are often limited to linear resonators or
oscillators with specific topologies, our definition is applicable to any
oscillator as a figure of merit for its amplitude stability.
It can be formulated rigorously and computed numerically from oscillator
equations.
With this definition, we calculate and analyze the \Q{} factors of several
oscillators of different types.
The results confirm that the proposed \Q{} formulation is a useful addition to
the characterization techniques for oscillators.
\end{abstract}

\section{\normalfont {\large Introduction}}\seclabel{introduction}

The concept of quality factor, or \Q{} factor, is one that is applicable in
many fields of engineering.
The \Q{} factor is a dimensionless parameter that describes how ``lossy'' an
oscillator is.
A higher \Q{} factor has many implications.
It is often taken as synonymous with the stability of the oscillator in both
amplitude and frequency, which then translates to better energy efficiency and
lower phase noise --- the ``quality'' is higher.
However, once we try to write down an exact formula for the \Q{} factor of an
oscillator, several confusions arise.

Firstly, \Q{} factor is often defined under the context of (usually second-order)
linear resonators, which are systems with damped oscillatory behaviours.  There
are several definitions.
One is the frequency-to-bandwidth ratio of the resonator:
\begin{equation}\eqnlabel{Qdef1}
	Q \stackrel{\mathrm{def}}{=} \frac{f_r}{\Delta f} = \frac{\omega_r}{\Delta \omega}.
\end{equation}

The formula implies that there is a Bode plot of the system with a resonance
frequency, thus is only meaningful for BIBO stable linear systems with
well-defined inputs and outputs.
It is not directly applicable to oscillators which are by-definition autonomous
and usually nonlinear.
Another definition for \Q{} factor is from the energy perspective:
\begin{equation}\eqnlabel{Qdef2}
	Q \stackrel{\mathrm{def}}{=} \frac{\text{Energy Stored}}{\text{Energy Dissipated per Cycle}}. 
\end{equation}

This assumes that there is damping in the oscillation, which is not true for
self-sustaining oscillators. 
There are other definitions that directly map Q to a parameter in the transfer
function, but they are limited to linear systems as well.

These definitions are rigorous and clear for linear resonators,
and they are widely-used in the design of integrated inductors and tuned
band-pass filters.
But they are not directly applicable to, and are often misused in the
characterization of nonlinear oscillators.
This leads to another common confusion ---  people often simply assume an
oscillator to have the same \Q{} factor as the resonator it is using inside.
For example, an LC oscillator is often said to have the same \Q{} as the linear
RLC circuit in it.
However, this is not true.
An obvious counterexample is that the use of a high-\Q{} resonator inside
does not guarantee a high-\Q{} oscillator; one has to be careful in the
design of the nonlinear circuit around it in order not to introduce too much
distortion or power loss.

For oscillators, one may extend the \Q{} definition in \eqnref{Qdef2} to undamped
oscillation.
This requires manually identifying energy storage components and integrating
energy dissipations.
It depends heavily on the topology of the oscillator and is only applicable to
several types of LC oscillators.
Another approach attempts to extend the frequency-domain characterization
technique in \eqnref{Qdef1} to feedback oscillators.
The oscillator is assumed to have a feedback block-diagram model with an
open-loop transfer function.
This implies linearization of the nonlinear components at the amplitude of
oscillation, analogous to the describing function analysis.
The transfer function is then assumed to have a resonance frequency, where
the slope of its phase with respect to frequency gives 
an ``open-loop \Q{}'' \cite{razavi1996phasenoise}.
The higher the open-loop \Q{}, the sharper the phase slope, and the more
resistant the closed-loop system to the variations in the frequency of
oscillation.
Similar open-loop \Q{} concept can also be defined using S-parameter transfer
functions \cite{randall2001Qdef}.
More recently, the relationship between this \Q{} definition and the phase
noise performances are discussed for oscillators with one- and two-port
passive linear networks \cite{ohira2005Qdef}.
These \Q{} definitions offer useful insights into an oscillator's frequency
stability. But for these frequency-domain, transfer-function-based techniques
to be applicable, specific circuit topology, sinusoidal waveforms and the
presence of a resonant subcircuit have to be assumed.

Another widely-used definition of \Q{} factor in oscillator measurements is
arguably adapted from \eqnref{Qdef1}.
It also uses a frequency-to-linewidth ratio:
\begin{equation}\eqnlabel{Qdef1}
	Q \stackrel{\mathrm{def}}{=} \frac{f}{\Delta f_{3dB}},
\end{equation}
where $f_{3dB}$ is the 3dB linewidth of the spectrum of the oscillator's
output.
This \Q{} factor can be conveniently measured using a spectrum analyzers, but
it is arguable if this quality factor is a property intrinsic to the
oscillator.
The linewidth depends on the noise sources in the circuit --- their types,
magnitudes, and locations.
This indicates that \Q{} factor defined in this way will be affected by
temperature and interferences.
And numerical calculation of this \Q{} factor won't be feasible without knowing
the noise model associated with the oscillator;
the characterization cannot be performed from just oscillator equations.

In this paper, we discuss a new definition of \Q{} factor that can be applied
to any oscillator.
It can be conveniently measured, rigorously formulated and numerically
computed.
It is defined from the energy perspective, in close analogy to the linear
resonator case in \eqnref{Qdef2}.
Instead of analyzing the total energy storage, we add (or subtract) a small
amount of extra energy to the oscillator by perturbation.
The oscillator will settle back to its amplitude-stable state eventually,
dissipating (or restoring) a fraction of the extra energy every cycle.
The speed in which the oscillator settles back indicates how amplitude-stable it
is, reflecting the energy efficiency of the oscillator; we can then define \Q{}
factor based on it.
We expand on this idea in \secref{definition}.

As it turns out, there is a better way to formulize and characterize this \Q{}
factor than directly applying perturbation and observing amplitude decay.
The energy-based intuition in \secref{definition} leads to a more
mathematically rigorous and general formulation of the proposed \Q{} factor,
which is closely related to Floquet theory, and can be calculated numerically
from oscillator equations.
We discuss the formulation and numerical characterization in more detail in 
\secref{LPTV} and \secref{characterization} respectively.
We then try the numerical techniques on several oscillators of different types,
including ring oscillators, LC oscillators, spin torque oscillators, and a
chemical reaction oscillator in \secref{examples}.
The examples confirm that the proposed \Q{} factor formulation matches the
intuition behind quality factor and offers a good measure of an oscillator's
amplitude settling behaviour.

In \secref{examples}, we also discuss the differences and relationship between
our new \Q{} factor definition and the other energy-based one in \eqnref{Qdef2}
in the context of LC oscillators.
We emphasize that the proposed \Q{} factor formulation is indeed different and
more suitable for characterizing amplitude stability.
In \secref{comparison}, we further summarize and compare definitions of \Q{}
for both linear resonators and nonlinear oscillators.
While the frequency- and energy-based definitions of \Q{} for second-order
linear resonators are equivalent, different \Q{} definitions for oscillators
are not.
The \Q{} formulation we propose is a useful addition to the figures of merit of
oscillators, especially in applications that require fast amplitude settling,
\eg, low distortion, amplitude-enhanced VCOs
\cite{deng2013CVCO,vannerson1974RCosc}, oscillator-based computation systems
\cite{WaRoUCNC2014PHLOGON,WaRoDAC2015MAPPforPHLOGON}, \etc.

\section{\normalfont {\large Intuition Behind the New \Q{} Definition}}
\seclabel{definition}

\figref{intuition} compares the waveforms of two oscillators.
Even without any definition, it is easy to tell that the LC oscillator should
have a higher \Q{}, since its settling behaviour is much slower, indicating
more stability in amplitude.
The definition of \Q{} should match this intuition.

\begin{figure}[htbp]
\centering{
    \epsfig{file=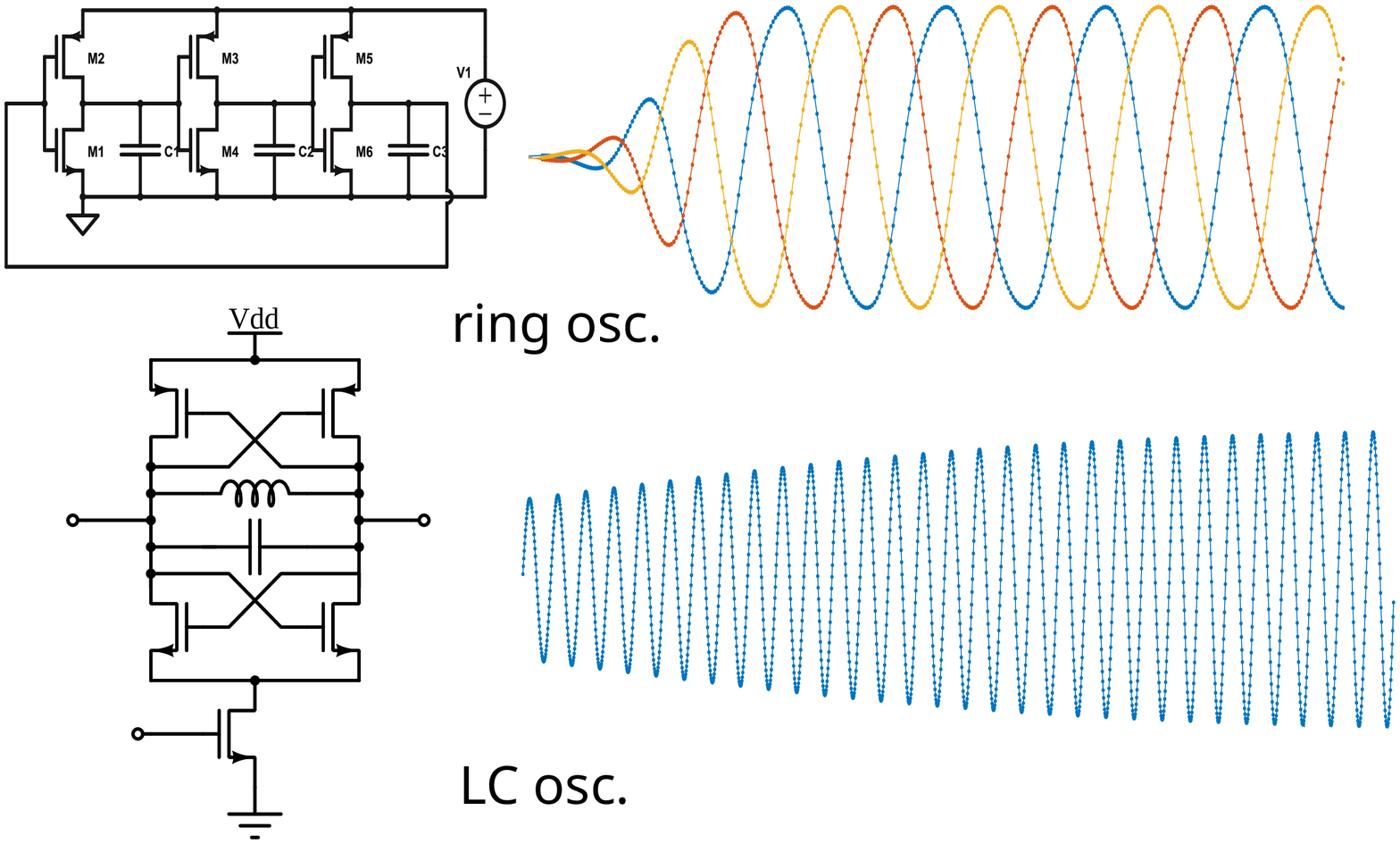,width=\linewidth}
}
\caption{Intuition behind \Q{} factor: LC oscillator (high-\Q{}) settles more
slowly in amplitude than ring oscillator (low-\Q{}).
	\figlabel{intuition}}
\end{figure}

We define \Q{} from the energy perspective, similar to \eqnref{Qdef2}.
Consider perturbing an 
oscillator with a small amount of extra energy; the oscillator settles back to
its stable orbit gradually.
We can then define the ratio between the extra energy applied and the energy
dissipated (or restored) every cycle as the \Q{} factor of the oscillator:
\begin{equation}\eqnlabel{Qdef}
	Q \stackrel{\mathrm{def}}{=}\frac{\text{Extra Energy Applied}}{\text{Extra Energy Dissipated per Cycle}}. 
\end{equation}

\begin{figure*}[htbp]
\centering{
    \epsfig{file=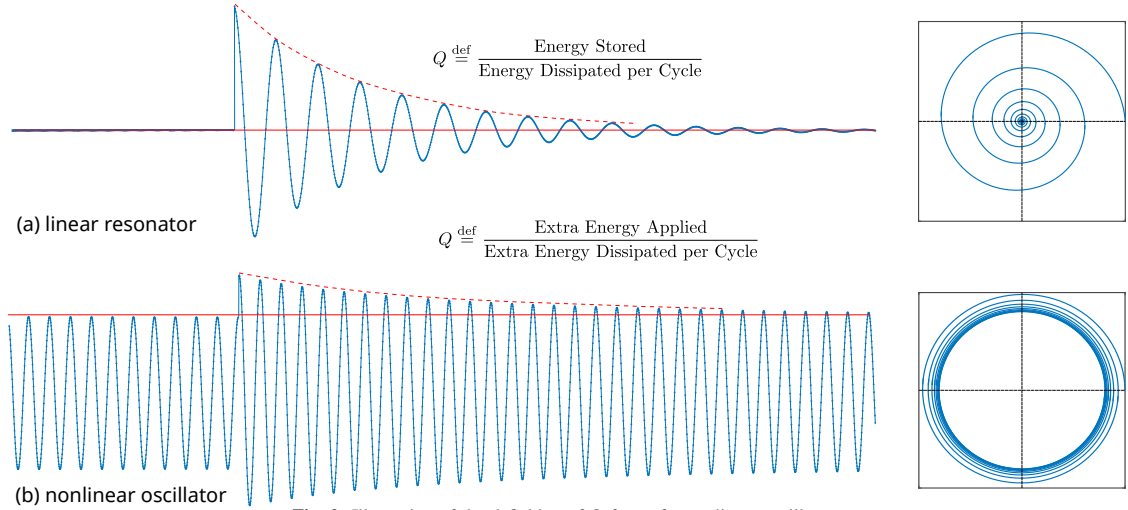,width=0.8\linewidth}
}
\caption{Illustration of the definition of Q factor for nonlinear
oscillators.
    \figlabel{Qfactor}}
\end{figure*}

The definition is illustrated in \figref{Qfactor} with both the time-domain
waveforms and phase plane plots.
It can be conveniently measured from either transient simulations or
experiments, without analyzing circuit topology or open-loop block diagrams.
The higher the \Q{} factor, the more slowly its amplitude responds to
perturbations, and the more stable the amplitude of the limit cycle.
These observations fit intuition.

As also illustrated in \figref{Qfactor}, the \Q{} definition for oscillators is
analogous to that for the linear resonators.
And the widely-used energy-based \Q{} formulation in \eqnref{Qdef2} for linear
systems is in fact just a special case of our definition in \eqnref{Qdef}, with
the amplitude-stable state being the zero state.

Note that the definition in \eqnref{Qdef} has several limitations when applied
in practice.
As the oscillator is nonlinear, different amount of the extra energy introduced
will have different results.
Ideally, to study the properties of the limit cycle, we would like the extra
energy to be as small as possible, which then makes measurement difficult when
there is noise or numerical error.
Also, the direction of the perturbation in the state space also affects the
settling speed.
Despite the limitations, the energy-based definition serves as the physical
intuition behind the more rigorous definition we discuss in \secref{LPTV}.


\section{\normalfont {\large Formulation of \Q{} from LPTV Theory}}
\seclabel{LPTV}

To study an oscillator's response under a small perturbation, we first model
the oscillator mathematically as a set of Differential Algebraic Equations
(DAEs) in the following form.
\begin{equation}\eqnlabel{oscDAE}
    \frac{d}{dt} \vec q (\vec x(t)) + \vec f(\vec x(t)) = \vec 0,
\end{equation}
where $\vec x(t) \in \mathbb{R}^n$ contains the time-varying unknowns, or states in the
oscillator system; $\vec q$ and $\vec f$ are vector functions with the same
size as $\vec x(t)$.

The oscillator system is autonomous, \ie, DAE \eqnref{oscDAE} has no
time-varying inputs, but the steady state response is time-varying and
periodic, \ie, there exists a non-constant response $\vec x_s(t)$ satisfying
\eqnref{oscDAE} and a time period $T > 0$, such that
\begin{equation}\eqnlabel{PSS}
\vec x_s(t) = \vec x_s(t+T).
\end{equation}

If the oscillator starts at $t=0$ with $\vec x_s(0)$, $\vec x_s(t)$ will be its
response.
Now we consider perturbing the initial condition by a small $\Delta \vec x_0$,
the response for $t \ge 0$ will satisfy
\begin{equation}\eqnlabel{perturbedDAE}
    \frac{d}{dt} \vec q (\vec x_s(t) + \Delta \vec x(t)) + \vec f(\vec x_s(t) +
    \Delta \vec x(t)) = \vec 0,
\end{equation}
with
\begin{equation}
    \Delta \vec x(0) = \Delta \vec x_0,
\end{equation}
where the time-varying vector $\Delta \vec x(t)$ captures the derivation of the
oscillator's response under perturbation from its steady state $\vec x_s(t)$.

If $\Delta \vec x(t)$ is always small with its magnitude decaying over time,
the trajectory $\vec x_s(t) + \Delta \vec x(t)$ is close to $\vec x_s(t)$.
Linearizing $\vec q$ and $\vec f$ along $\vec x_s(t)$ results in
a Linear Periodic Time Varying (LPTV) system as in \eqnref{LPTV}, which gives a
good approximation to the original system in \eqnref{perturbedDAE}.
\begin{equation}\eqnlabel{LPTV}
    \frac{d}{dt} \mathbf{C}(t) \cdot \Delta \vec x(t) + \mathbf{G}(t) \cdot \Delta
    \vec x(t) = \vec 0,
\end{equation}
where $\mathbf{C}(t)$ and $\mathbf{G}(t)$ are time-varying, $T$-periodic
matrices.
\begin{equation}
    \mathbf{C}(t) = \left. \frac{d \vec q}{d \vec x} \right|_{\vec x_s(t)}, ~~
    \mathbf{G}(t) = \left. \frac{d \vec f}{d \vec x} \right|_{\vec x_s(t)}.
\end{equation}

As it turns out, the assumption that $\Delta \vec x(t)$ decays with time is not
always true for arbitrary perturbations; there are constraints on $\Delta \vec
x_0$ for the approximation to hold.
We will concretize these constraints as we proceed with our discussion.

In order to analyze the LPTV system in \eqnref{LPTV}, we consider solving for
the fundamental matrix associated with it.
The fundamental matrix of a LPTV system is the time-varying matrix solution
$\mathbf{X}(t) \in \mathbb{R}^{n\times n}$ to the following matrix DAE.
\begin{equation}\eqnlabel{Xt}
    \frac{d}{dt} \mathbf{C}(t) \cdot \mathbf{X}(t) + \mathbf{G}(t) \cdot
    \mathbf{X}(t) = \vec 0,
\end{equation}
with
\begin{equation}\eqnlabel{X0}
    \mathbf{X}(0) = \mathbf{I_n}.
\end{equation}

$\mathbf{X}(t)$ at a given time $t$ can be calculated numerically through
time-integration.
Once $\mathbf{X}(t)$ is calculated, for any small initial perturbation $\Delta
\vec x_0$, $\Delta \vec x(t)$ is simply given as 
\begin{equation}
    \Delta \vec x(t)  = \mathbf{X}(t) \cdot \Delta \vec x_0,~~ t \ge 0.
\end{equation}

In particular, $\Delta \vec x(T)  = \mathbf{X}(T) \cdot \Delta \vec x_0$
captures the changes to the initial perturbation $\Delta \vec x_0$ after one cycle of
oscillation.
According to Floquet theory \cite{Farkas94}, there exists a time-invariant matrix
$\mathbf{F}$ such that 

\begin{equation}
    \mathbf{X}(t)  = \mathbf{U}(t) \cdot e^{\mathbf{F} \cdot t} \cdot \mathbf{V}^*(t),
\end{equation}
where $\mathbf{U}(t)$ and $\mathbf{V}(t)$ are $T$-periodic, non-singular
matrices.

The eigenvalues of matrix $\mathbf{F}$, namely $\{\mu_k\}$, are known as the
{\it Floquet exponents}.
The eigenvalues of matrix $\mathbf{X}(T)$, namely $\{\lambda_k\}$, can then be expressed
as
\begin{equation}
   \lambda_k = e^{\mu_k \cdot T}.
\end{equation}

$\{\lambda_k\}$ are known as {\it characteristic multipliers}.
Their moduli $\{|\lambda_k|\}$ can be calculated from the real parts of
$\{\mu_k\}$, which are also called {\it Lyapunov exponents} \cite{Farkas94}.
This $\{\lambda_k\}$ spectrum of matrix $\mathbf{X}(T)$ gives us
comprehensive information on the changes of $\Delta \vec x_0$ after one cycle
of oscillation.

It is provable that the oscillation is self-sustaining if and only if there exists
a {\it characteristic multiplier} exactly equal to 1 \cite{DeMeRoTCAS2000}.
This means that part of the initial perturbation $\Delta x_0$ will never
disappear; it represents the phase change of the oscillation after
perturbation.
Without loss of generality, we assume $\lambda_1 = 1$, and then arrange the rest
of $\{\lambda_k\}$ in descending order of their moduli.
Among them, the one with the largest modulus is $\lambda_2$.
If $|\lambda_k| < 1$ for $k \ge 2$, and $\Delta x_0$ doesn't contain any component
aligned with the eigenvector corresponding to $\lambda_1$, the initial
perturbation will decay with time.
And the speed of decay is characterized by its slowest moment, which decreases
to $|\lambda_2|$ of its size after each cycle.
If we assume that after \Q{} cycles the size is no larger than $5\%$ of its
initial value, then
\begin{equation}\eqnlabel{Qformulation}
   |\lambda_2|^{Q} \leq 0.05 \implies Q \ge \log_{|\lambda_2|} 0.05.
\end{equation}

\Q{} factor can then be estimated as the value of $\log_{|\lambda_2|} 0.05$, or
the integer closest to it.

Note that for capturing the amplitude settling behaviour precisely, we will
need the whole spectrum of matrix $\mathbf{X}(T)$, \ie, $\{\lambda_k\}$.
The response of a LPTV system, just like that of a LTI system, consists of the
superposition of components associated with all {\it multipliers}.
However, if we were to derive a single scalar figure of merit to represent this
complex settling behaviour, it is not unreasonable to consider only the slowest
moment, characterized by $\lambda_2$.

The formulation of \Q{} is based on the exponential decay of $\Delta \vec x(t)$
after time $T$.
We argue that it is indeed based on the definition of \eqnref{Qdef}.
The energy associated with a DAE is often formulated as
\begin{equation}\eqnlabel{Hamiltonian}
   E(\vec x) \propto \vec x^T \mathbf{H} \vec x,
\end{equation}
where $\mathbf{H}$ is a symmetric matrix commonly known as the Hamiltonian of
the system. Its exact formula varies from system to system.
From \eqnref{Hamiltonian}, we have
\begin{equation}
   \Delta E(\Delta \vec x, \vec x) = E(\vec x + \Delta \vec x) - E(\vec x) \propto \vec x^T \mathbf{H} \cdot \Delta \vec x,
\end{equation}
where $\mathbf{H}$ is a fixed matrix, $\vec x$ is the trajectory of oscillation
--- both are independent of $\Delta \vec x$. Therefore, the energy decay
$\Delta E$ is equivalent to the settling of perturbation $\Delta x$,
characterized by the same set of {\it characteristic multipliers}.

\section{\normalfont {\large Numerical Characterization}}
\seclabel{characterization}

From \secref{LPTV}, the \Q{} formulation in \eqnref{Qformulation} can be
computed numerically.
The procedures are summarized as follows:
\begin{enumerate}
\item Compute the large-signal periodic steady state solution $\vec x_s(t)$
	  numerically, often through the shooting method \cite{kundertwhite},
      or Harmonic Balance \cite{BaBiCh92}.
\item Linearize functions $\vec q(\vec x)$ and $\vec f(\vec x)$ along $\vec
	  x_s(t)$, calculate time-varying Jacobian matrices $\mathbf{C}(t)$ and
	  $\mathbf{G}(t)$, $0 \leq t \leq T$, store them as tabulated data
	  \footnote{$\mathbf{C}(t)$ and $\mathbf{G}(t)$ can be pre-computed
	  and are often stored at the same time as the periodic steady state
      calculation.}.
\item Solve for $\mathbf{X}(T)$ through numerical time-integration of
	  the matrix DAE \eqnref{Xt} from its initial condition \eqnref{X0}.
\item Perform eigenanalysis on $\mathbf{X}(T)$ to obtain $\lambda_2$;
	  calculate \Q{} factor with \eqnref{Qformulation}.
\end{enumerate}

We note that there are well-developed numerical techniques for each step in the
calculation of \Q{}. 
Many implementations are already available today in open-source and commercial
simulators for oscillator phase noise analysis.
We point out that these implementations can be easily adapted to the analysis
of amplitude stability of oscillators.

We also note that, for improved efficiency, full eigenanalysis on
$\mathbf{X}(T)$ is not necessary.
Since the calculation of \Q{} only requires $\lambda_2$ and we know that
$\lambda_1=1$, power method can be easily modified to compute only $\lambda_2$.

Our implementation is based on MAPP: the Berkeley Model and Algorithm
Prototyping Platform \cite{WaAaWuYaRoCICC2015MAPP,MAPPwebsite}. We plan to
release it as open-source software under the GNU General Public License.

\section{\normalfont {\large Examples}}\seclabel{examples}

After explaining the intuition, formulation and calculation of the proposed new
\Q{} definition in the previous sections, in this section, we compute and
analyze the \Q{} factors of several oscillators of different types. 

\subsection{Ring Oscillators}

\subsubsection{Analytical Results from Ideal Ring Oscillator}

The simplest ideal ring oscillator has 3 stages, each of which consists of
a memoryless ideal inverter followed by an RC delay circuit.
It can be modelled using the input voltages ($v_1$, $v_2$ and $v_3$) of the
three stages:
\begin{equation}
    \dot{v_1} = \frac{f(v_3) - v_1}{\tau},~
    \dot{v_2} = \frac{f(v_1) - v_2}{\tau},~
    \dot{v_3} = \frac{f(v_2) - v_3}{\tau},
\end{equation}
where $\tau$ is the time constant of the RC circuit at each stage; $f(v)$
represents the I/O relationship of the ideal inverter:
\begin{equation}
    f(v) = \begin{cases}
    -1, & \text{if } v \ge 0,\\
    +1, & \text{otherwise.}
    \end{cases}
\end{equation}

The reason for considering this ideal ring oscillator is that, despite its
simplicity, it is well suited for estimating both qualitative and quantitative
features of ring oscillators \cite{RoVLSID05}, and its fundamental matrix
solution is available analytically \cite{RoVLSID05}.

The three eigenvalues of $\mathbf{X}(T)$ are $\lambda_1 = 1$,
$\lambda_2 = \phi^{-6}$,
$\lambda_3 = \phi^{-12}$,
where $\phi = (\sqrt{5}+1)/2 \approx 1.618$ is known as the golden ratio.

In this case, 
$\lambda_2 = \phi^{-6} \approx 0.0557$.
Therefore, $Q \approx 1.1$.
This matches intuition, as the ideal ring oscillator dissipates all of its
energy after every cycle of oscillation, thus restoring to its stable amplitude
very quickly.

\subsubsection{Numerical Results from CMOS Ring Oscillator}
To test the numerical characterization techniques in \secref{characterization},
we simulate a more realistic
3-stage CMOS ring oscillator with BSIMv3.3 models.
The circuit schematic is the same as in \figref{intuition}; the waveform is
also similar to the one in it.
As we can see, the waveform looks almost sinusoidal, without the sharp turns in
the ideal ring oscillator case.
But numerical computation reveals that $\lambda_2 \approx 0.067$.
\Q{} is still close to 1, even though the waveform is less distorted.
The oscillator's amplitude stability information is indeed embedded in its
equations.
And our \Q{} formulation captures it truthfully.

\subsection{LC Oscillators}

\subsubsection{High-\Q{} and Low-Q{} Negative Resistance LC Oscillators}
Here we consider a simple negative-resistance LC oscillator circuit shown in
\figref{negLC}.
Different nonlinearities in $f(v)$ result in different \Q{} factors.
Intuitively, as $f(v)+\frac{1}{R}$ gets ``flatter'', the oscillator will appear
more like an LC tank with no resistance, thus the quality \Q{} gets higher.

\begin{figure}[htbp]
\centering{
	\epsfig{file=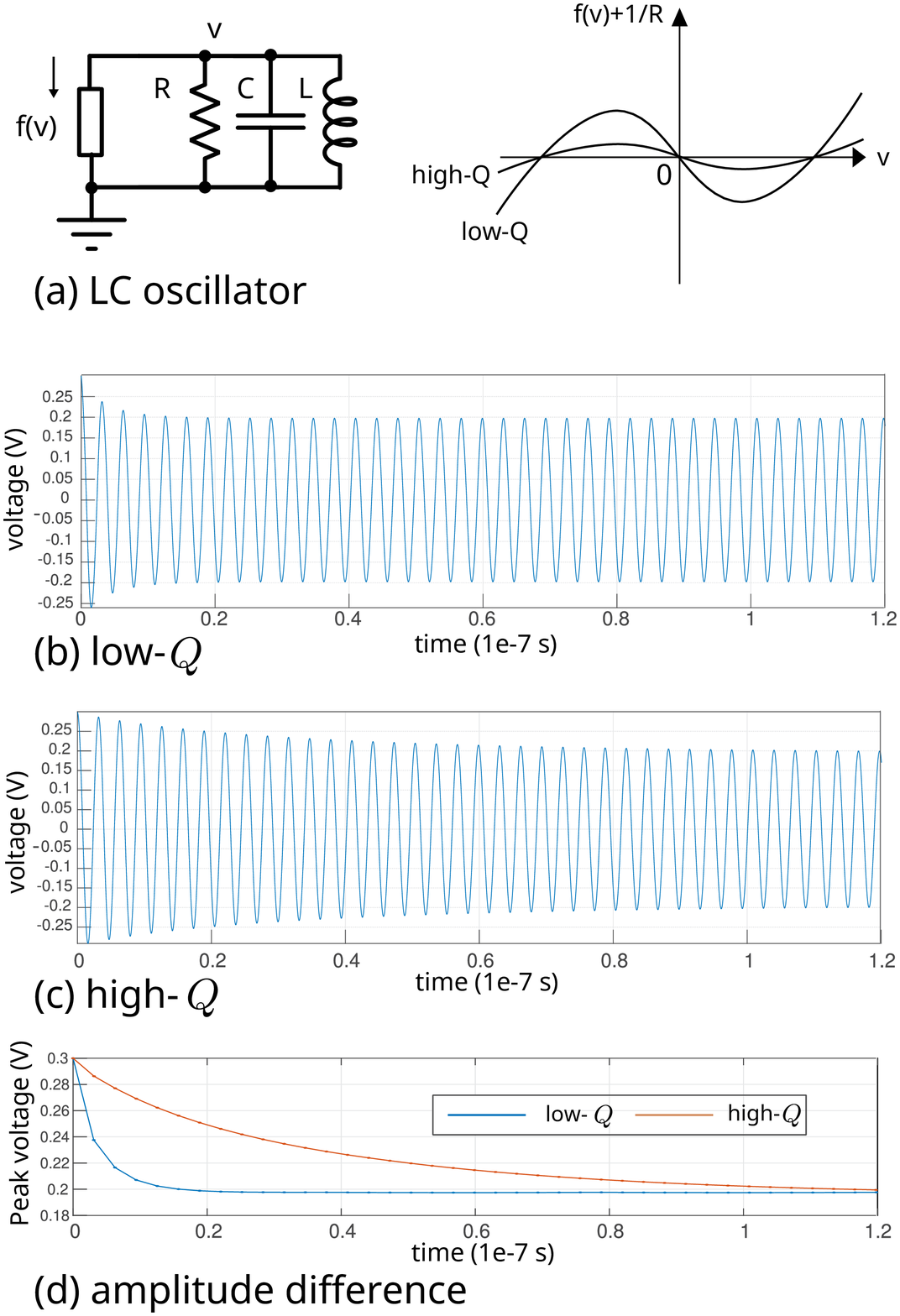,width=\linewidth}
}
\caption{(a) A simple negative-resistance LC oscillator. Different
choice of f(v) will result in different \Q{} factor of the oscillator.
(b-d) the changes of amplitudes after perturbations for low-\Q{} and high-\Q{} LC
oscillators.
	\figlabel{negLC}}
\end{figure}

We choose $L=0.5nH$, $C=0.5nF$, $f(v) = K \cdot (v - \tanh(1.01\cdot v))$ for
the LC oscillator.
We choose $K$ values as 1 and 20, the former results in a high-\Q{} oscillator,
the latter low-\Q{}.
Numerical \Q{} factor characterization results in $\lambda_2 \approx 0.54$ for
the former, and $\lambda_2 \approx 0.94$ for the latter,
indicating \Q{} factors of $4.8$ and $48$ respectively.

The transient simulation results in \figref{negLC} confirm that the high-\Q{}
one indeed settles much more slowly in amplitude.

\subsubsection{\Q{} Factor and Energy Dissipation}
In the circuit in \figref{negLC}, the inductor and capacitor are ideal devices
that do not dissipate any energy. 
When the circuit exhibits self-sustaining oscillation, the energy dissipated
by the resistor and that generated by the negative resistance component should
balance out in every cycle.
If we assume the voltage waveform at the only non-ground node in the circuit
to be sinusoidal as $v(t) = V_{\max}\sin(\omega t)$,
the two energies per cycle can be plotted as in \figref{disc_func}.

\begin{figure}[htbp]
\centering{
	\epsfig{file=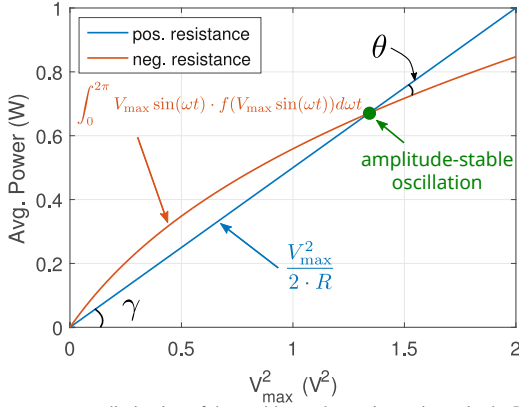,width=0.75\linewidth}
}
\caption{Average power dissipation of the positive and negative resistors in
the LC oscillator in \figref{negLC}. Their intersection indicates
self-sustaining oscillation. The angle $\theta$ characterizes how
amplitude-stable the oscillation is.
	\figlabel{disc_func}}
\end{figure}

In \figref{disc_func} we have chosen a slightly different set of parameters
with $f(v) = K \cdot (v - \tanh(2\cdot v))$ such that the curves are more
clearly separated.
From \figref{disc_func}, we observe that between $V_{\max} = 0$ and the
oscillation point, the negative resistance device generates more energy per
cycle than that dissipated by the positive resistor.
Therefore, the oscillation cannot be stable and the amplitude will build up.
On the other hand, when $V_{\max}$ is beyond the oscillation point, the
amplitude of oscillation will decrease.
Therefore, only at the oscillation point will the two energies balance out and
the amplitude be stable.
The angle $\theta$ then characterizes how fast amplitude settles back to this
equilibrium.
More specifically, if $V^2_{\max}$ changes by $\delta$, $(\tan(\gamma+\theta) -
\tan(\gamma))\cdot \delta$ captures how much extra energy will be dissipated or
restored in one cycle of oscillation.
This is to say, $1 - (\tan(\gamma+\theta) - \tan(\gamma))$ is proportional to
$|\lambda_2|$.
Given $\gamma$, the larger the value of $\theta$, the smaller the value of
$|\lambda_2|$, and the lower the \Q{} factor.
This offers a graphical interpretation of the \Q{} factor for this particular
type of LC oscillator.

In comparison, we consider the \Q{} definition in \eqnref{Qdef2}.
In this LC circuit, the definition is applicable since we can identify the
energy storage and dissipation components.
At the oscillation point, the two average powers in \figref{disc_func} are
equal.
Therefore, the angle $\gamma$, determined only by the positive resistor, will
characterize the energy dissipation in the \Q{} definition.
This \Q{} definition doesn't use any information from the nonlinearity in the
circuit, nor does it provide any information related to the oscillation point,
\eg, the angle $\theta$.
It captures the energy dissipation correctly, but it doesn't characterize the
amplitude stability of the oscillator.

In conclusion, these two definitions correspond to different properties in the
energy plot; they characterize different attributes of the oscillator.

\subsection{BJT Phase-shift Oscillator}

The \Q{} factors of ring oscillators and LC oscillators are straightforward to
understand; our calculations match the intuitions behind them.
But unlike conventional \Q{} definitions, our formulation works for general
nonlinear oscillators beyond these two well-studied types.

A typical BJT phase-shift oscillator as shown in \figref{BJT-phaseshift-osc}
consists of an inverting BJT amplifier and a phase-shift RC network that feeds
the delayed output as the input to the amplifier.
It does not require an inductor to oscillate, and there is no band-pass
linear resonator in the oscillator;
defining the \Q{} factor by identifying energy storage devices becomes less
relevant.
But our \Q{} formulation can still be applied.
$\lambda_2$ for the circuit in \figref{BJT-phaseshift-osc} is 0.9081,
corresponding to a \Q{} factor of 31.
The results show that although there is no inductor, the oscillator does not
behave like a ring oscillator, which settles to the limit cycle very fast.
Instead, the oscillation builds up after several cycles.
This observation for the \Q{} factor characterization can be confirmed with
transient simulations as well, \eg, in \figref{BJT-phase-shift-osc-tran}.
In this case, for transient simulation to show the amplitude settling
behaviour, initial guesses that are far away enough from the limit cycle have
to be used.
And as the system has multiple dimensions, randomly chosen initial guesses will
show different amplitude settling speed; they won't capture the worst case that
defines the \Q{} factor.
In comparison, the LPTV-based characterization is more rigorous and more
accurate in measuring the stability of the limit cycle.

\begin{figure}[htbp]
\centering{
	\epsfig{file=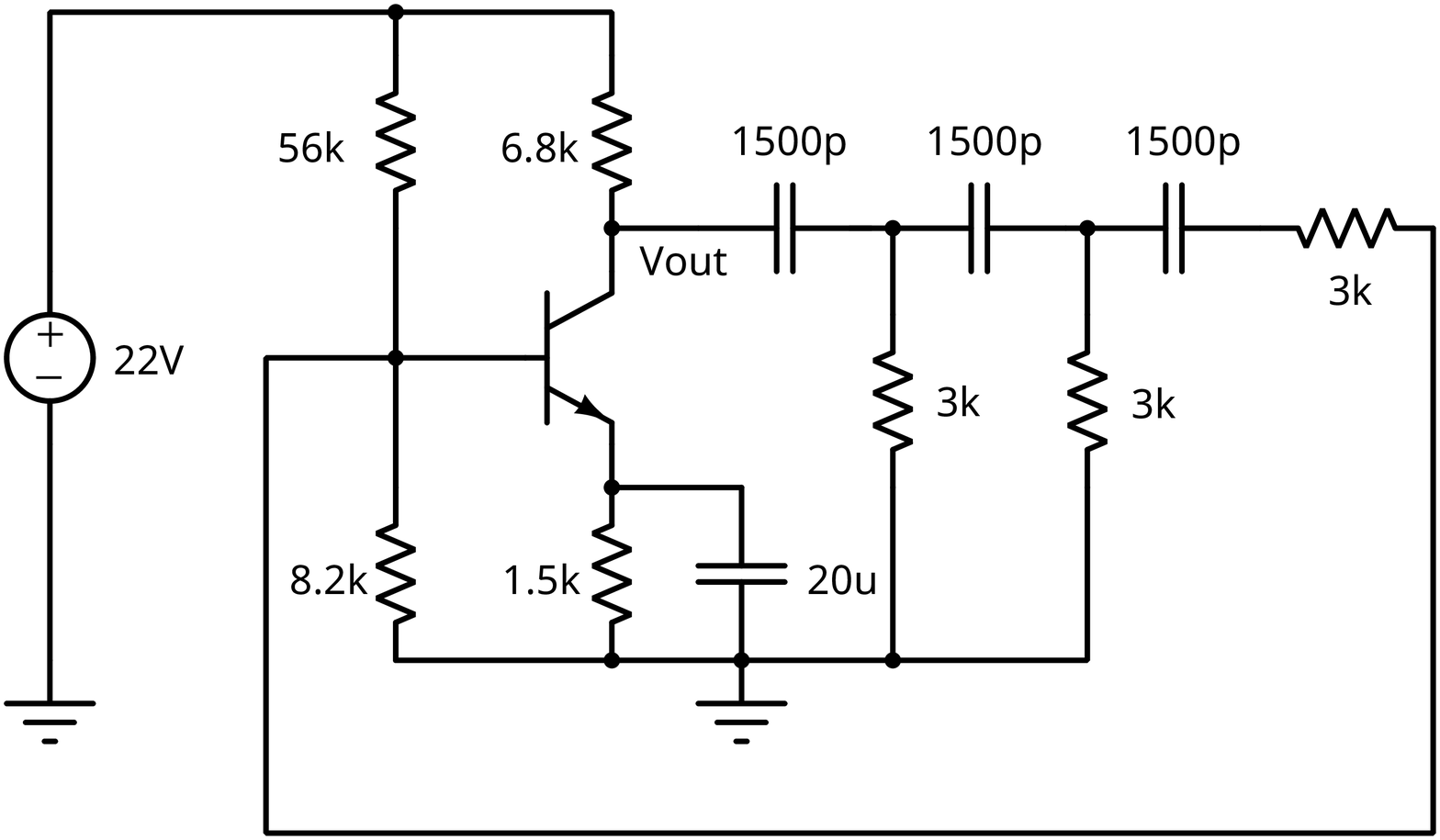,width=0.7\linewidth}
}
\caption{Schematic of a BJT phase-shift oscillator. \figlabel{BJT-phaseshift-osc}}
\end{figure}

\begin{figure}[htbp]
\centering{
	\epsfig{file=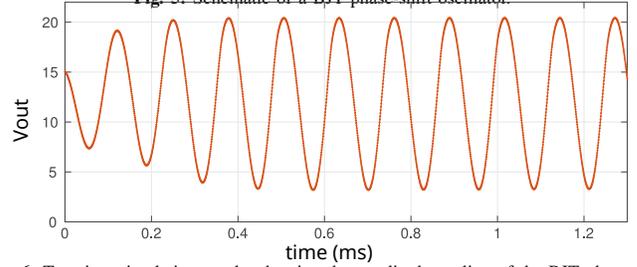,width=0.9\linewidth}
}
\caption{Transient simulation results showing the amplitude settling
of the BJT phase-shift oscillator. \figlabel{BJT-phase-shift-osc-tran}}
\end{figure}

\subsection{Spin Torque Nano-Oscillator (STNO): LLG Equation}

Magnetization in a STNO, often represented with a 3D vector $\mathbf{M}$, can
vary internally in response to torques $\mathbf{H_\mathrm{ext}}$.
The rotation of $\mathbf{M}$ is predicted by the Landau-Lifshitz-Gilbert (LLG)
equation (adapted and simplified from \cite{camsari2015spinlib}):
\begin{equation}\eqnlabel{LLG}
    \tau
	\frac{d\mathbf{M}}{dt}= - \mathbf{M} \times \mathbf{H_\mathrm{eff}}
     - \alpha \mathbf{M} \times \left(\mathbf{M} \times
       \mathbf{H_{\mathrm{eff}}}\right)
     - \alpha \mathbf{M} \times \left(\mathbf{M} \times
       \mathbf{I_{\mathrm{s}}}\right),
\end{equation}
where
\begin{equation}\eqnlabel{LLG2}
       \mathbf{H_{\mathrm{eff}}} = 
       \mathbf{H_{\mathrm{int}}} + 
       \mathbf{H_{\mathrm{ext}}} = 
       [K_x,~K_y,~K_z]^T \cdot \mathbf{M} + 
       \mathbf{H_{\mathrm{ext}}}.
\end{equation}

{\small
\begin{table}[ht]
    \centering
    \begin{tabular}{|m{0.3\linewidth}|m{0.3\linewidth}|}
        \hline
         Parameter Name & Value \\
        \hline
        \hline
         $\tau$ & 1e-9 \\
        \hline
		 $\alpha$  &  0.02 \\
        \hline
		 $K_x$  &  -10 \\
        \hline
		 $K_y$  &  0 \\
        \hline
		 $K_z$  &  1 \\
        \hline
		 $\mathbf{I_{\mathrm{s}}}$ & $[0, 0, -0.6mA]^T$  \\
        \hline
		 $\mathbf{H_{\mathrm{ext}}}$ & $[0, 0, 2]^T$  \\
        \hline
    \end{tabular}
	\vspace{0.5em}
	\caption{Parameters and inputs for the LLG equation of a
STNO.}\tablabel{parameters}
\end{table}
}

Both inputs $\mathbf{H_\mathrm{ext}}$ and $\mathbf{I_\mathrm{s}}$ are 3D
vectors.
When they have appropriate values, the magnetization vector $\mathbf{M}$ can
begin self-sustaining oscillation, which generates an oscillating voltage
across the device.
An example set of parameters that can generate oscillating $\mathbf{M}$ is
given in \tabref{parameters}.

\begin{figure}[htbp]
\centering{
	\epsfig{file=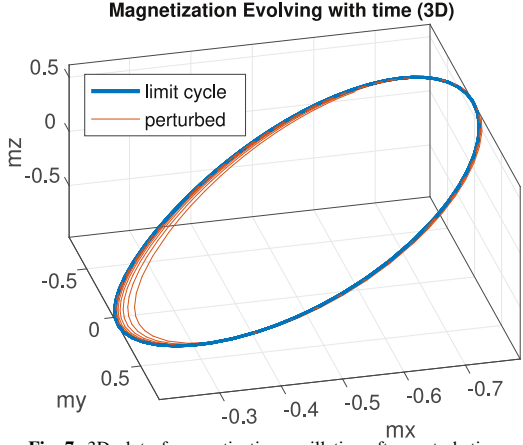,width=0.75\linewidth}
}
\caption{3D plot of magnetization oscillation after perturbation.
 \figlabel{LLG_perturbed_3D}}
\end{figure}

\ignore{
\begin{figure}[htbp]
\centering{
	\epsfig{file=./figures/LLG_perturbed_mx.eps,width=\linewidth}
}
\caption{Magnetization oscillation along the x axis after perturbation.
\figlabel{LLG_perturbed_mx}}
\end{figure}
}

There are no inductors or capacitors in the system; identifying energy storage
is not straightforward.
It is not clear what the noise sources are in the system.\footnote{It
is often assumed that noise is another $\mathbf{H_\mathrm{noise}}$ magnetic
field term added to \eqnref{LLG}. But study showing good match of this
modelling assumption and measurements is still lacking.}
The only way for characterizing \Q{} is by fabricating the oscillator and
measuring it with a spectrum analyzer; conventional \Q{} characterizations
cannot be carried out with system equations in \eqnref{LLG} and \eqnref{LLG2}.
In comparison, our formulation of \Q{} still applies.
With the set of parameter listed in \tabref{parameters}, we simulate a STNO
and show results in \figref{LLG_perturbed_3D}. 
\Q{} characterization shows that $\lambda_2 \approx 0.9712$, $Q \approx 103$.
We note that the slow amplitude settling behaviour is due to the nonlinear
resonator inside the oscillator, which is passive and under-damped just like an
RLC tank.
Because of the nonlinearity, conventional Bode-based analysis cannot be
performed. 
But the limit cycle is relatively stable compared with ring oscillators; it
behaves more like LC oscillator.
Our \Q{} characterization matches this intuition.
These results show the generality of our approach --- the \Q{} formulation is
not limited to oscillators with linear resonators; it is not limited to
electrical oscillators either.

%

\subsection{A Chemical Reaction Oscillator}

Here we consider a set of chemical reactions that constitute a chemical
oscillator.
\begin{eqnarray}
	A + B &\xrightarrow{[k]}& 2B \\
	B + C &\xrightarrow{[k]}& 2C \\
	C + A &\xrightarrow{[k]}& 2A
\end{eqnarray}

The oscillator is nonlinear, with the concentrations of the three species as
its states. 
In this case, it is not clear how to identify the energy loss in the system, or
to make any analogy to commonly-known electrical oscillators; conventional \Q{}
factor definitions and characterization techniques are not applicable.
In comparison, our \Q{} factor formulation still works and can capture
interesting properties of this oscillator.

\begin{figure}[htbp]
\centering{
	\epsfig{file=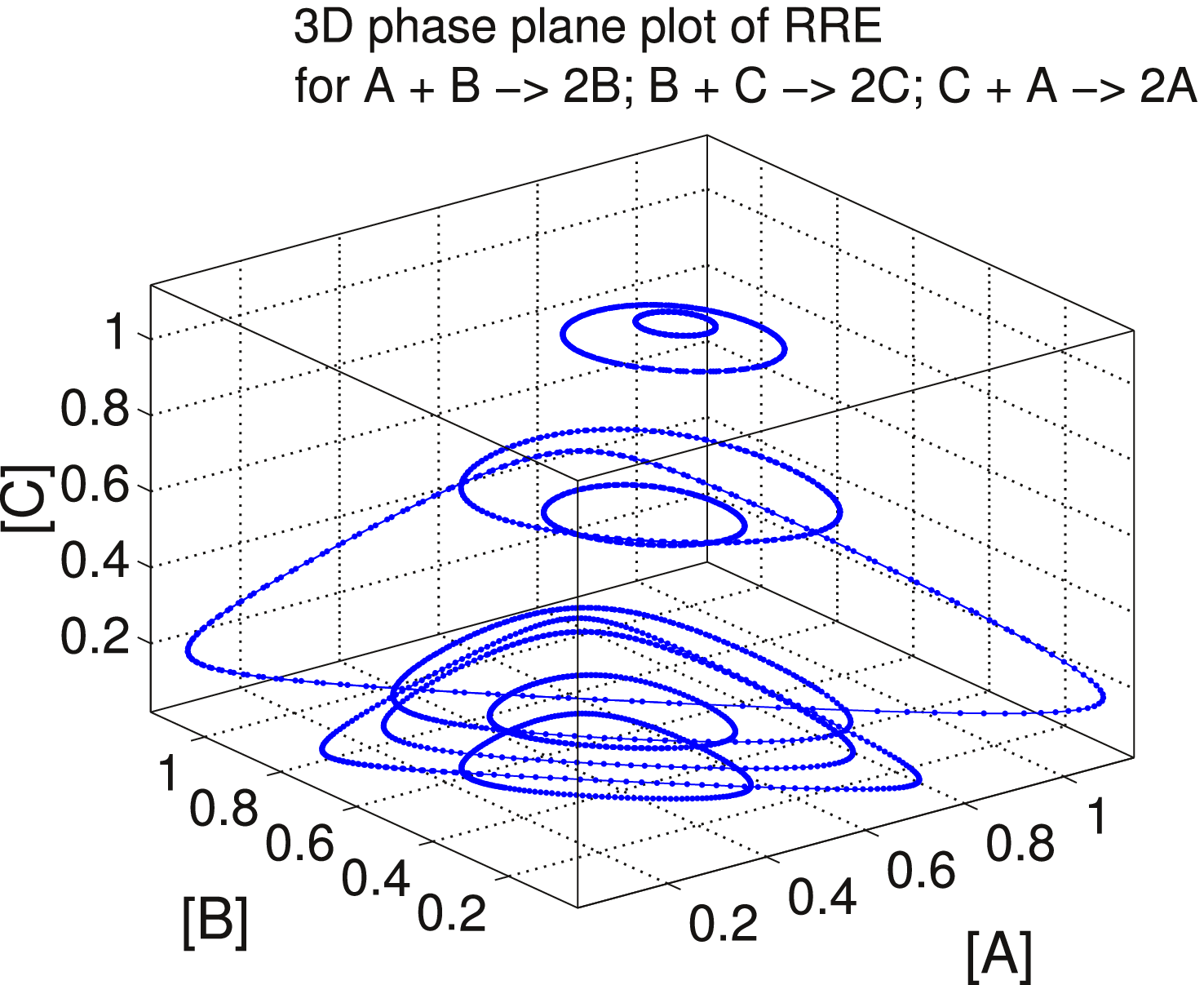,width=0.7\linewidth}
}
\caption{3D phase plane plot of a chemical reaction oscillator with different
		 random initial concentrations.
	\figlabel{RRE}}
\end{figure}

Numerical calculation of $\lambda_2$ returns an almost perfect $1$, up to 14
digits, indicating an almost infinite \Q{} factor.
Put in other words, the fundamental matrix $\mathbf{X}(T)$ has two eigenvalues
equal to $1$, indicating that both the phase and amplitude of this oscillator
are unstable.
This indeed matches intuition.
Starting from different random initial concentrations of the three species,
oscillations can be observed to happen on planes governed by the conservation
law of this chemical system.
A perturbation to the states can knock the limit cycle to a nearby plane, and
the extra energy is never dissipated.
These can be confirmed with transient simulation results in \figref{RRE}.

\section{\normalfont {\large Comparison of Different Q factor Definitions}}\seclabel{comparison}

For second-order linear resonators, \Q{} factor is well defined, with two
definitions --- one based on the frequency-domain representation,
the other from the energy perspective.
We have introduced them in \secref{introduction}. 
Here we formulate them again, with names $Q_{l1}$ and $Q_{l2}$.

The frequency-domain definition $Q_{l1}$ is based on the Bode plot of the
system:
\begin{equation}\eqnlabel{Ql1}
	Q_{l1} \stackrel{\mathrm{def}}{=} \frac{f_r}{\Delta f} = \frac{\omega_r}{\Delta \omega},
\end{equation}
where $\Delta f$ is the 3dB bandwidth of the Bode plot.

The energy-based definition $Q_{l2}$ is
\begin{equation}\eqnlabel{Ql2}
	Q_{l2} \stackrel{\mathrm{def}}{=} \frac{\text{Energy Stored}}{\text{Energy Dissipated per Cycle}}. 
\end{equation}

For second-order linear systems, these two definitions are equivalent and are
often used interchangeably.
To see this, we consider the transfer function of a second-order resonator:
\begin{equation}\eqnlabel{Linear1}
H(s) = \frac{Y(s)}{X(s)} = \frac{2\zeta\omega_0 \cdot s}{s^2 + 2\zeta\omega_0 \cdot s + \omega_0^2},
\end{equation}
where $X(s)$ and $Y(s)$ are the Laplace transform of the input and output
respectively.
When input $x(t)$ is an impulse, the output is 
\begin{equation}\eqnlabel{Linear2}
y(t) = A \cdot e^{- \zeta\cdot \omega_0 \cdot t} \cdot \cos(\omega_0\cdot \sqrt{1-\zeta^2}\cdot t + \phi),
\end{equation}
where $A$ and $\phi$ depend on the initial condition of the system and the size
of the impulse.
The resonance frequency is $\omega_r = \omega_0\cdot \sqrt{1-\zeta^2}$ rad/s,
or $f_r = \frac{\omega_0\cdot \sqrt{1-\zeta^2}}{2\pi}$ Hz.

The magnitude of the Bode plot is
\begin{equation}
|H(j\omega)| = \frac{2\zeta\omega_0 \omega}{|\omega_0^2-\omega^2 + j2\zeta\omega_0 \omega|}.
\end{equation}

At the peak of the Bode plot, $s = j\omega_r$. When the system is under-damped,
$\zeta$ is small, $\omega_r\approx \omega_0$. Therefore, 
\begin{equation}
|H(j\omega_r)| \approx 1.
\end{equation}

The 3dB bandwidth corresponds to two frequencies $\omega_{1,2}$ where
$|H(j\omega_{1,2})| \approx 1/\sqrt{2}$.
This indicates
\begin{equation}
|\omega_0^2-\omega_{1,2}^2| = |2\zeta\omega_0 \omega_{1,2}|,
\end{equation}
which results in
\begin{equation}
\omega_{1,2} = \omega_0\cdot (\sqrt{\zeta^2+1} \pm \zeta).
\end{equation}

The 3dB bandwidth is 
\begin{equation}
\Delta\omega = \omega_2 - \omega_1  = 2\zeta \omega_0.
\end{equation}

$Q_{l1}$ can then be written as 
\begin{equation}
Q_{l1} \stackrel{\mathrm{def}}{=} \frac{\omega_r}{\Delta \omega} = 
\frac{\omega_0\cdot \sqrt{1-\zeta^2}}{2\zeta \omega_0} \approx \frac{\sqrt{1-\zeta^2}}{2\zeta}.
\end{equation}

To formulate $Q_{l2}$, from \eqnref{Linear2}, the amplitude of the damped
oscillation decays exponentially with time constant $\tau = \frac{1}{\zeta\cdot
\omega_0}$.
Therefore, after one cycle $T = \frac{2\pi}{\omega_0\cdot \sqrt{1-\zeta^2}}$s,
the amplitude decays to $e^{-T/\tau}$ of its original value, corresponding to
an energy decay to $e^{-2T/\tau}$.
\begin{equation}
	Q_{l2} = \frac{1}{1 - e^{-2T/\tau}} = \frac{1}{1 - e^{-\frac{4\pi\zeta}{\sqrt{1-\zeta^2}}}}.
\end{equation}

When the system is under-damped, $\zeta$ is small,
\begin{equation}
    1 - e^{-\frac{4\pi\zeta}{\sqrt{1-\zeta^2}}} \approx \frac{4\pi\zeta}{\sqrt{1-\zeta^2}}.
\end{equation}

Therefore,
\begin{equation}\eqnlabel{equiv}
	Q_{l2} \approx \frac{\sqrt{1-\zeta^2}}{4\pi\zeta} = \frac{1}{2\pi}Q_{l1}.
\end{equation}

The two definitions for the \Q{} factor of second-order linear systems are
indeed equivalent, with only a constant factor between them.
This conclusion is not new, but the discussion here is useful in understanding
the confusion behind the definitions of \Q{} for nonlinear oscillators.
It is arguably because of this elegant equivalence of \Q{} definitions for
second-order linear systems that we would naturally but erroneously expect that
there exists a single \Q{} definition for nonlinear systems as well, leading to
some of the misconceptions and misuse of \Q{} factors for oscillators.

In fact, as has been shown recently \cite{tsakmakidis2017QfactorScience}, even
for linear systems, once they are not required to consist of all passive
elements, the approximate equivalence in \eqnref{equiv} can be violated ---
energy dissipation and bandwidth can indeed be decoupled.
When it comes to nonlinear oscillators, there is no reason for us to expect a
single \Q{} that captures all the oscillator's properties; the different
available definitions warrant discussions.

For oscillators, neither $Q_{l1}$ nor $Q_{l2}$ applies directly.
Instead, \Q{} factor is defined vaguely following the intuitions behind them.
It can be done in several ways.

As mentioned in \secref{introduction}, one definition is frequency based, using the
3dB bandwidth of the spectrum of the output of the oscillator.
\begin{equation}\eqnlabel{Qn1}
	Q_{n1} \stackrel{\mathrm{def}}{=} \frac{f}{\Delta f_{3dB}}.
\end{equation}

While the spectrum of the oscillating output signal indeed looks like a Bode
plot, it is fundamentally different.
Bode plot portraits the transfer function from the input to the output of a
stable system; it is intrinsic to the system and can be calculated from system
equations.
For an oscillator, the sharpness of the peak observed in a spectrum analyzer
depends on the noise sources present in the system.
And it changes with temperature and noise level.
So while $Q_{n1}$ characterizes the frequency stability of the oscillator, it
is a different quantity from $Q_{l1}$ for linear systems.

The energy-based definition of \Q{} factor, \ie, $Q_{l2}$, can be extended for
oscillators in more than one ways.
The commonly-used definition requires calculating the total energy stored in
the oscillator and the energy that is dissipated per cycle:
\begin{equation}
	Q_{n2} \stackrel{\mathrm{def}}{=} \frac{\text{Energy Stored}}{\text{Energy Dissipated per Cycle}}. 
\end{equation}

This definition is useful for analyzing the energy efficiency of the
oscillator.
But it is only meaningful when there are energy storage devices inside the
oscillator.
For oscillators that don't contain LC tanks, especially those outside the
electrical domain, such as chemical reaction and spin-torque oscillators, it is
difficult to define this $Q_{n2}$ rigorously.

Although $Q_{n2}$ is widely used, it represents only one way of interpreting
the energy-based $Q_{l2}$ in the context of nonlinear oscillators.
In this paper, we have proposed a new definition of \Q{}, namely $Q_{n3}$, that
is based on perturbation on the limit cycle.
\begin{equation}
	Q_{n3} \stackrel{\mathrm{def}}{=}\frac{\text{Extra Energy Applied}}{\text{Extra Energy Dissipated per Cycle}}. 
\end{equation}

From our discussion in \secref{LPTV}, we can argue that this $Q_{n3}$ is more
closely analogous to $Q_{l2}$, as the extra amplitude and energy indeed decay
to the limit cycle exponentially.
The $Q_{l2}$ for linear systems becomes a special case of $Q_{n3}$ where the
``limit cycle'' is the zero state.
$Q_{n3}$ can be more rigorously formulated using the Lyapnov exponents of the
system, which are analogous to the exponent for the decay of magnitude in
second-order linear systems in \eqnref{Linear2}.
It is a quantity intrinsic to the oscillator, and can be calculated
analytically or numerically from system equations.

The three \Q{} definitions --- $Q_{n1}$, $Q_{n2}$ and $Q_{n3}$ --- characterize
an oscillator's frequency stability, energy efficiency, amplitude stability
respectively.
However, unlike in the linear resonator case, there is no direct equivalence
between them.
In fact, for some oscillators, they can be distinctly different.
For example, a ring oscillator dissipates almost all the stored energy during
each cycle; its amplitude also restores very quickly.
Therefore, it has low $Q_{n2}$ and $Q_{n3}$.
But it can still have a large $Q_{n1}$ based on the frequency-domain
definition.
Yet we rarely see reports of high-\Q{} ring oscillators, because the use of
$Q_{n2}$ is often assumed.
Similar to ring oscillators, spin-torque oscillators require a DC current to
sustain oscillation; they keep draining energy from the DC supply while
retaining little energy in the oscillation.
This means $Q_{n2}$ should be close to 0.
But high-\Q{} spin-torque oscillators have been reported with \Q{} factors over
3000 \cite{maehara2014highQstno}; they are analyzed based on the definition of
$Q_{n1}$ instead.
This inconsistent use of terminology can create confusions for researchers if
they don't have a clear understanding of the multiple \Q{} factor definitions
available for oscillators.
For LC oscillators, as we show in \secref{examples}, with a fixed $Q_{n2}$,
depending on the design of the nonlinearity, $Q_{n3}$ can vary, reflecting the
stability of the limit cycle of the oscillator.  In this case, $Q_{n3}$ is
arguably more useful for design.

In summary, while $Q_{l1}$, $Q_{l2}$ are equivalent, $Q_{n1}$, $Q_{n2}$ and
$Q_{n3}$ are not; these quality factors characterize three separate
``qualities'' of nonlinear oscillators.
Among them, the $Q_{n3}$ we discuss in this paper is a new one that can
be defined rigorously, characterized mathematically, calculated numerically
given oscillator equations, and it has close relationship with linear systems'
\Q{} factor. It has not been studied before.

\section*{\normalfont {\large Conclusion}}

In this paper, we described our ideas on defining \Q{} factor of a nonlinear
oscillator by perturbing it and assessing its amplitude stability.
We discussed the mathematical formulation and numerical calculation procedures
of the proposed \Q{} definition.
Results on different oscillators have validated its usefulness in the
characterization of oscillators.
In our further research,
we plan to study the effects of the proposed \Q{} factor on the phase
noise performances of oscillators.
We also intend to extend the proposed concept and techniques to systems that
are not a single oscillator, including multi-tone systems and oscillators with
ancillary non-oscillatory circuitry.

\renewcommand{\baselinestretch}{0.8}
\let\em=\it
\bibliographystyle{unsrt}
\bibliography{stringdefs,jr,von-Neumann-jr,PHLOGON-jr,tianshi}

\begin{thebibliography}{10}

\bibitem{razavi1996phasenoise}
B.~Razavi.
\newblock {A study of phase noise in CMOS oscillators}.
\newblock {\em Solid-State Circuits, IEEE Journal of}, 31(3):331--343, 1996.

\bibitem{randall2001Qdef}
M.~Randall and T.~Hock.
\newblock {General oscillator characterization using linear open-loop
  S-parameters}.
\newblock {\em Microwave Theory and Techniques, IEEE Transactions on},
  49(6):1094--1100, 2001.

\bibitem{ohira2005Qdef}
T.~Ohira.
\newblock {Rigorous Q-factor formulation for one-and two-port passive linear
  networks from an oscillator noise spectrum viewpoint}.
\newblock {\em Circuits and Systems II: Express Briefs, IEEE Transactions on},
  52(12):846--850, 2005.

\bibitem{deng2013CVCO}
W.~Deng, K.~Okada, and A.~Matsuzawa.
\newblock {Class-C VCO with amplitude feedback loop for robust start-up and
  enhanced oscillation swing}.
\newblock {\em Solid-State Circuits, IEEE Journal of}, 48(2):429--440, 2013.

\bibitem{vannerson1974RCosc}
{E. Vannerson and K. Smith}.
\newblock {Fast amplitude stabilization of an RC oscillator}.
\newblock {\em {IEEE Journal of Solid-State Circuits}}, 9(4):176--179, 1974.

\bibitem{WaRoUCNC2014PHLOGON}
{T. Wang and J. Roychowdhury}.
\newblock {PHLOGON: PHase-based LOGic using Oscillatory Nanosystems}.
\newblock In {\em Proc. UCNC}, LNCS sublibrary: Theoretical computer science
  and general issues. Springer, July 2014.
\newblock \putDOI{\href{http://dx.doi.org/10.1007/978-3-319-08123-6_29}{DOI
  link.}}

\bibitem{WaRoDAC2015MAPPforPHLOGON}
{T. Wang and J. Roychowdhury}.
\newblock {Design Tools for Oscillator-based Computing Systems}.
\newblock In {\em Proc.\/ IEEE DAC}, pages 188:1--188:6, 2015.
\newblock \putDOI{\href{http://dx.doi.org/10.1145/2744769.2744818}{DOI link.}}

\bibitem{Farkas94}
M.~Farkas.
\newblock {\em Periodic Motions}.
\newblock Springer-Verlag, New York, 1994.

\bibitem{DeMeRoTCAS2000}
A.~Demir, A.~Mehrotra, and J.~Roychowdhury.
\newblock {Phase Noise in Oscillators: a Unifying Theory and Numerical Methods
  for Characterization}.
\newblock {\em IEEE Trans.~Ckts.~Syst.~--~I: Fund.~Th.~Appl.}, 47:655--674, May
  2000.
\newblock \putDOI{\href{http://dx.doi.org/10.1109/81.847872}{DOI link.}}

\bibitem{kundertwhite}
K.S. Kundert, J.K. White, and A.~Sangiovanni-Vincentelli.
\newblock {\em Steady-state methods for simulating analog and microwave
  circuits}.
\newblock Kluwer Academic Publishers, 1990.

\bibitem{BaBiCh92}
J.W. Bandler, R.M. Biernacki, and S.H. Chen.
\newblock {Harmonic balance simulation and optimization of nonlinear circuits}.
\newblock In {\em Proc. IEEE ISCAS}, volume~1, pages 85--88, May 1990.

\bibitem{WaAaWuYaRoCICC2015MAPP}
{T. Wang, K. Aadithya, B. Wu, J. Yao, and J. Roychowdhury}.
\newblock {MAPP: The Berkeley Model and Algorithm Prototyping Platform}.
\newblock In {\em Proc. IEEE CICC}, pages 461--464, September 2015.
\newblock \putDOI{\href{http://dx.doi.org/10.1109/CICC.2015.7338431}{DOI
  link.}}

\bibitem{MAPPwebsite}
{MAPP: The Berkeley Model and Algorithm Prototyping Platform}.
\newblock Web site:
  \href{https://github.com/jaijeet/MAPP/wiki}{https://github.com/jaijeet/MAPP/wiki}.

\bibitem{RoVLSID05}
J.~Roychowdhury.
\newblock {Exact Analytical Equations for Predicting Nonlinear Phase Errors and
  Jitter in Ring Oscillators}.
\newblock In {\em Proc. IEEE Conf. VLSI Design}, January 2005.

\bibitem{camsari2015spinlib}
K.~Camsari, S.~Ganguly, and S.~Datta.
\newblock {Modular approach to spintronics}.
\newblock {\em Scientific reports}, 5.

\bibitem{tsakmakidis2017QfactorScience}
{K.L, Tsakmakidis, L. Shen, S.A. Schulz, X. Zheng, J. Upham, X. Deng, H. Altug,
  A.F. Vakakis, R.W. Boyd}.
\newblock {Breaking Lorentz reciprocity to overcome the time-bandwidth limit in
  physics and engineering}.
\newblock {\em Science}, 356(6344):1260--1264, 2017.

\bibitem{maehara2014highQstno}
{H. Maehara, H. Kubota, Y. Suzuki, T. Seki, K. Nishimura, Y. Nagamine, K.
  Tsunekawa, A. Fukushima, H. Arai, T. Taniguchi, and others}.
\newblock {High Q factor over 3000 due to out-of-plane precession in
  nano-contact spin-torque oscillator based on magnetic tunnel junctions}.
\newblock {\em Applied Physics Express}, 7(2):023003, 2014.

\end{thebibliography}

\end{document}